\documentclass[final,3p,times,twocolumn]{elsarticle}
 \biboptions{comma,sort&compress}
\usepackage{here}
 \usepackage{graphicx}
  \usepackage{epsfig}
\def\nin{\noindent}
\def\beq{\begin{equation}}
\def\eeq{\end{equation}}
\def\bea{\begin{eqnarray}}
\def\eea{\end{eqnarray}}

\journal{Nuc. Phys. (Proc. Suppl.)}

\begin{document}

\begin{frontmatter}



\title{Diphoton production at Next-to-Next-to-Leading-Order}

% \author[label1]{Leandro Cieri\corref{cor1}}
 \author[label1]{Leandro Cieri}
  \address[label1]{INFN, Sezione di Firenze, 
\\
Via G.Sansone 1, I-50019 Sesto Fiorentino, Florence, Italy.}
%\cortext[cor1]{Speaker}
\ead{cieri@fi.infn.it}

%\author{}

%\address{}

\begin{abstract}
\noindent
We consider direct diphoton
%photon pair 
production in hadron collisions. We compute the next-to-next-to-leading
order (NNLO) QCD radiative corrections at the fully-differential level.
% and
% we show how the NNLO corrections are essential
%to undertand the main background concerning recents Higgs boson searches.
Our calculation uses the $q_T$ subtraction formalism and it is implemented
in a parton level Monte Carlo program, which allows the user to apply arbitrary kinematical cuts on the
final-state photons and the associated jet activity, and to compute the
corresponding distributions in the form of bin histograms. We present selected
numerical results related to Higgs boson searches at the LHC and the Tevatron, and we show how
the NNLO corrections to diphoton production are relevant to understand the main background
of the decay channel $H\rightarrow\gamma \gamma$ of the Higgs boson $H$.
\end{abstract}


\end{frontmatter}

% \linenumbers

\section{Introduction}
%\label{}
\nin
%%%%%%%%%%%%
Diphoton production is a relevant process
in hadron collider physics.
It is both a classical signal within the Standard Model (SM)
and an important background for Higgs boson and new-physics
searches.
The origin of the Electroweak symmetry breaking is currently
being
investigated at the LHC by searching for the Higgs boson and
eventually studying its properties. 
Recent results in the search for the SM Higgs Boson at the LHC indicates the observation of a new particle~\cite{cha:2012gu,aad:2012gk}, which is a neutral boson with a mass $M\sim 125$~GeV. 
%Therefore the motivation in searches and studies, based in the fact that the Higgs boson mass must be low ($114$~GeV$< m_H < 130$~GeV), is renovated by this spectacular new observation. And thus, as in previous searches and studies, the preferred search mode involves Higgs boson production via gluon fusion followed by the rare decay into a pair of photons.
In this spectacular new observation, as well as in previous searches and studies, the preferred search mode involves Higgs boson production via gluon fusion followed by the rare decay into a pair of photons. 
Therefore, it is essential to count on an accurate theoretical description of
the various kinematical distributions associated to the production of pairs of
prompt photons with large invariant mass.
%Such task requires detailed computations of radiative corrections.

\vspace*{0.2cm}

\nin
We are interested in the process $pp \rightarrow \gamma \gamma X$,
which, at the lowest order in perturbative QCD, occurs \textit{via} 
the quark annihilation subprocess $q\bar{q} \rightarrow \gamma \gamma$. The QCD corrections at 
the next-to-leading order (NLO) in the strong coupling $\alpha_{\mathrm{S}}$ involve the  
quark annihilation channel and a new partonic channel, \textit{via} the
subprocess $qg \rightarrow \gamma \gamma q$. These corrections have been
computed and implemented in the fully-differential Monte Carlo codes
\texttt{DIPHOX}~\cite{Binoth:1999qq}, \texttt{2gammaMC} \cite{Bern:2002jx} and 
\texttt{MCFM}~\cite{Campbell:2011bn}. A calculation that includes the effects of
 transverse-momentum resummation is implemented in 
\texttt{RESBOS}~\cite{Balazs:2007hr}. At the next-to-next-to-leading order 
(NNLO), 
the $gg$ channel starts to contribute,
and
the large gluon--gluon luminosity makes this channel sizeable. 
Part of the contribution from this channel,
the so called {\it box contribution}, was computed long ago \cite{Dicus:1987fk} and 
its size turns out to be comparable  
to the lowest-order result.

\vspace*{0.2cm}

\nin
Besides their {\it direct} production from the hard subprocess, photons can also
arise from fragmentation subprocesses of QCD partons. The computation of
fragmentation subprocesses requires (poorly known)
non-perturbative information, in the form of 
parton  
fragmentation functions of the photon.
The complete NLO single- and double-fragmentation contributions are implemented in \texttt{DIPHOX}~\cite{Binoth:1999qq}. The effect of the fragmentation contributions 
is sizeably reduced by the photon isolation criteria that are 
necessarily
applied in hadron collider experiments to suppress the very large irreducible
background (e.g., photons that are faked by jets or produced by hadron decays). 
The standard cone isolation and the `smooth' cone isolation proposed
by Frixione \cite{Frixione:1998jh} are two of these criteria. The standard cone
isolation is easily implemented in experiments, but it only suppresses a fraction of the 
fragmentation contribution.
The smooth cone isolation (formally) eliminates the entire fragmentation 
contribution, but its experimental implementation is still under study~\cite{CERN-note}. However, it 
is important to anticipate (work to appear) that in some kinematical regions (e.g for Higgs boson searches) QCD calculations that apply the standard cone and the Frixione
isolation criteria give basically very similar quantitative results\footnote{The use of the same parameters
in both criteria is understood.}.
%%%%%%%%%%%%
\section{Diphoton production at NNLO}
\nin
We consider the inclusive hard-scattering reaction
\begin{equation}
\label{one}
h_1+h_2\to \gamma\gamma +X \;\;,
\end{equation}
where the collision of the two hadrons, $h_1$ and $h_2$,
produces the diphoton system 
$F \equiv \gamma\gamma$ with high invariant mass $M_{\gamma \gamma}$.
The evaluation of the
NNLO corrections to this process requires the knowledge 
of the corresponding partonic scattering amplitudes
with $X=2$~partons (at the tree level \cite{Barger:1989yd}), $X=1$~parton (up 
to the one-loop level \cite{Bern:1994fz})
and no additional parton (up to the two-loop level \cite{Anastasiou:2002zn})
in the final state.
The implementation of the separate scattering amplitudes in a complete
NNLO (numerical) calculation is severely complicated by 
the presence of infrared (IR) divergences that occur at intermediate stages. 
The $q_T$ subtraction formalism \cite{Catani:2007vq} is a method that handles
and cancels these unphysical IR divergences up to the NNLO.
The formalism applies to generic hadron collision processes that involve
hard-scattering production of a colourless high-mass system $F$.
 Within that framework~\cite{Catani:2007vq}, the corresponding cross section is written as:
\bea
\label{main}
d{\sigma}^{F}_{(N)NLO}={\cal H}^{F}_{(N)NLO}\otimes d{\sigma}^{F}_{LO}
&+& \big[ d{\sigma}^{F+{\rm jets}}_{(N)LO}\nonumber \\
&-& d{\sigma}^{CT}_{(N)LO}\big]\;\;,
\eea
where $d{\sigma}^{F+{\rm jets}}_{(N)LO}$ represents the cross section for the
production of the system $F$ plus jets at (N)LO accuracy~\footnote{In the case of
diphoton production, the NLO calculation of 
$d{\sigma}^{\gamma\gamma+{\rm jets}}_{NLO}$ was performed in 
Ref.\cite{DelDuca:2003uz}.}, and
$d{\sigma}^{CT}_{(N)LO}$ is a (IR subtraction) counterterm whose explicit expression \cite{Bozzi:2005wk}
is obtained from the resummation program of the logarithmically-enhanced
contributions to $q_T$ distributions. 
The `coefficient' ${\cal H}^{F}_{(N)NLO}$, which also compensates for the subtraction
of $d{\sigma}^{CT}_{(N)LO}$,
corresponds to the (N)NLO truncation of the process-dependent perturbative function
\begin{equation}
{\cal H}^{F}=1+\frac{\alpha_{\mathrm{S}}}{\pi}\,
{\cal H}^{F(1)}+\left(\frac{\alpha_{\mathrm{S}}}{\pi}\right)^2
{\cal H}^{F(2)}+ \dots \;\;.
\end{equation}
The NLO calculation  of $d{\sigma}^{F}$ 
requires the knowledge
of ${\cal H}^{F(1)}$, and the NNLO calculation also requires ${\cal H}^{F(2)}$. The general 
structure of ${\cal H}^{F(1)}$
is explicitly known~\cite{deFlorian:2000pr}; exploiting the explicit results of ${\cal H}^{F(2)}$ for Higgs
\cite{Catani:2007vq,Catani:2011kr} and vector boson \cite{Catani:2009sm} 
production, we have generalized the process-independent relation of Ref.~\cite{deFlorian:2000pr} to 
the calculation of the NNLO coefficient 
${\cal H}^{F(2)}$.
%${\cal H}^{F(1)}$ is directly obtained from the process-dependent scattering
%amplitudes by using a process-independent relation.
%Exploiting the explicit results of ${\cal H}^{F(2)}$ for Higgs
%\cite{Catani:2007vq,Catani:2011kr} and vector boson \cite{Catani:2009sm} 
%production,
%we have generalized
%the process-independent relation of 
%Ref.~\cite{deFlorian:2000pr} to the calculation of the NNLO coefficient 
%${\cal H}^{F(2)}$ (this general result is presented in a forthcoming paper).
%
%Using this relation and the relevant scattering amplitudes
%\cite{Barger:1989yd,Bern:1994fz,Anastasiou:2002zn}, we have explicitly
%determined ${\cal H}^{F(2)}$ for diphoton production.
%
%
%%%%%%%%%%%%%%%%
\section{Quantitative results}
\nin
We have performed our fully-differential NNLO calculation~\cite{Catani:2011qz} of diphoton production
according to Eq.~(\ref{main}).
The NNLO computation is encoded
in a parton level
Monte Carlo program, in which
we can implement arbitrary IR safe cuts on the final-state
photons and the associated jet activity. 
%The present formulation of the $q_T$ subtraction formalism \cite{Catani:2007vq}
%is restricted to the production of colourless systems $F$ and, hence, it does not 
%treat parton fragmentation subprocesses (here $F$ includes one or two coloured
%partons that fragment).
We concentrate on the direct production of diphotons, and 
we rely on the smooth cone isolation criterion~\cite{Frixione:1998jh}. Considering a cone of radius $r=\sqrt{(\Delta \eta)^2+(\Delta \phi)^2}$ around
each photon, we require the total amount of hadronic (partonic) transverse energy $E_T$ 
inside the cone to be smaller than $E^{had}_{T\, max}(r)$,
\begin{equation}
\label{eq:etmax}
E_{T}<E^{had}_{T\, max}(r) \equiv  E_{T~max} \left(\frac{1-\cos r}{1- \cos R}\right)^{n}\,,
\end{equation}
where $E_{T~max}$ can be a fixed value or a fraction of the transverse momentum of the photon ($E_{T~max}=\epsilon_\gamma \,p_T^\gamma$, with $0<\epsilon_\gamma\leq 1$); the isolation criterion $E_T < E^{had}_{T\, max}(r)$ has to be fulfilled for all cones with $r\leq R$. We use the MSTW 2008~\cite{Martin:2009iq} sets of parton distributions, with densities and $\alpha_{\mathrm{S}}$ evaluated at each corresponding order,
and we consider $N_f=5$ massless quarks/antiquarks and gluons in 
the initial state. The default
renormalization ($\mu_R$) and factorization ($\mu_F$) scales are set to the value
of the invariant mass of the diphoton system,
$\mu_R=\mu_F = M_{\gamma\gamma}$. The QED coupling constant $\alpha$ is fixed to $\alpha=1/137$.
%%
%
%\begin{table}[htbp]
%\begin{center}
%\begin{tabular}{|c|c|c|c|}
%\hline
%$\sigma$ (fb)& LO & NLO & NNLO\\
%%%\hline
%\hline
%$\mu_F=\mu_R=M_{\gamma\gamma}/2$ & $5045 \pm 1$ & $26581\pm 23$ & $45588 \pm 97$\\
%\hline
%$\mu_F=\mu_R=M_{\gamma\gamma}$ & $5712\pm 2$ & $ 26402\pm 25$ & $ 43315 \pm 54$ \\
%\hline
%$\mu_F=\mu_R=2M_{\gamma\gamma}$ & $6319\pm 2$ & $26045\pm 24$ & $ 41794 \pm 77$\\
%\hline
%\end{tabular}
%\end{center}
%\caption{{\em Cross sections for $pp\to \gamma\gamma+X$ at the LHC
%($\sqrt{s}=14$~{\rm TeV}). The applied cuts are described in the text.}}
%%\label{tab:lhc}
%\end{table}
%
%
%\vspace{0.4cm}
%%%%%%%%%%%%%%%%%%%%%%%
\begin{figure}[hbt] 
\centerline{\includegraphics[width=7.3cm]{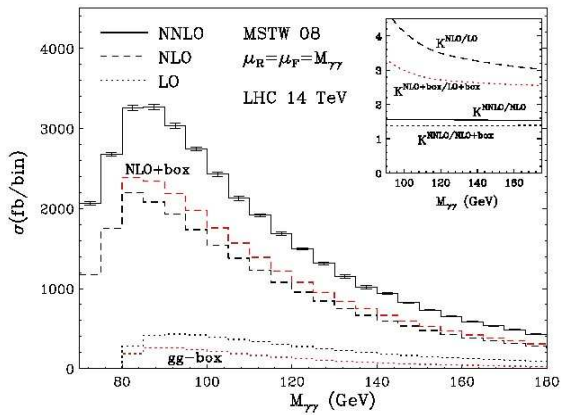}}
%{\epsfig{figure=mpsi2mc.eps,height=70mm}}
\caption{\scriptsize Invariant mass distribution of the photon pair
at the LHC 
($\sqrt{s}=14$~{\rm TeV}): LO (dots), NLO (dashes) and NNLO (solid) results. We
also present the results of the box and NLO+box contributions. The inset plot
shows the 
corresponding {\rm K}-factors.}
\label{fig:mass} 
\end{figure} 
\nin
%%%%%%%%%%%%%%%%%%%%%%%

\vspace*{0.5cm}

\nin
To present some quantitative results, we consider diphoton production
at the LHC ($\sqrt{s}=14$~TeV). We apply typical kinematical cuts used by ATLAS and CMS 
Collaborations in their Higgs boson search studies. We
require the harder and the softer photon to have transverse momenta $p_T^{\rm harder}\geq
40$~GeV and $p_T^{\rm softer}\geq 25$~GeV, respectively.
The rapidity of both photons is restricted to $|y_\gamma| \leq 2.5$, and
the invariant mass of the 
diphoton system is constrained to 
lie in the range $20 \,{\rm GeV}\leq M_{\gamma\gamma} \leq 250\,{\rm GeV}$.
The isolation parameters
are set to the values $\epsilon_\gamma=0.5$, $n=1$ and $R=0.4$. 
%all the numerical results presented in this Letter.
%In Table~\ref{tab:lhc}, we report the results
%of the accepted cross section at LO, NLO and NNLO.
%We have fixed $\mu_F=\mu_R=\mu$ and we have considered three values of 
%$\mu/M_{\gamma\gamma}$ ($\mu/M_{\gamma\gamma}=1/2,1,2$).
%The numerical errors estimate the statistical uncertainty of the Monte Carlo integration.
%%
%
%
Performing the QCD computation, we observe \cite{Catani:2011qz} that 
the value of the cross section remarkably increases with the perturbative order
of the calculation. This increase is mostly due to the use of {\it very
asymmetric} (unbalanced) cuts on the photon transverse momenta. At the LO,
kinematics implies that the two photons are produced with equal transverse
momentum and, thus, both photons should have $p_T^{\gamma}\geq 40$~GeV. 
At higher orders, the final-state radiation of additional partons opens a new
region of the phase space, 
where $40$~GeV $\geq p_T^{\rm softer}\geq 25$~GeV. Since photons can copiously be
produced with small transverse momentum~\cite{Catani:2011qz}, the cross section receives a sizeable
contribution from the enlarged phase space region. This effect is further
enhanced by the opening of a new large-luminosity partonic channel at each
subsequent perturbative 
order. 
%For example, at NLO the $qg$ channel accounts for about 80\% of the
%increase of the cross section.
%Therefore, it is not unexpected that a naive analysis of scale dependence% (as
%presented in Table~\ref{tab:lhc}) 
%underestimates
%the size of the higher-order corrections.
%
%
In Fig.~\ref{fig:mass} we show the LO, NLO and NNLO invariant mass
distributions at the default scales. 
We also plot the gluonic {\it box
contribution} (computed with NNLO parton distributions) and its sum with the full NLO result.
 The inset plot shows the K-factors defined as 
the ratio of the cross sections at two subsequent perturbative orders.
We note that ${\rm K}^{NNLO/NLO}$ is sensibly smaller than ${\rm K}^{NLO/LO}$,
and this fact indicates 
an improvement in the convergence of the perturbative expansion.
%%%%%%%%%%%%%%%%%%%%%%%
\begin{figure}[hbt] 
\centerline{\includegraphics[width=7.3cm,height=5.5cm]{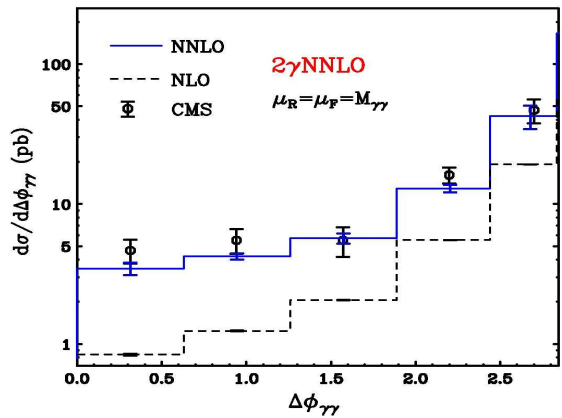}}
%{\epsfig{figure=mpsi2mc.eps,height=70mm}}
\caption{\scriptsize Diphoton 
cross section as a function of the azimuthal separation of
  the two photons. Data from CMS~\cite{Chatrchyan:2011qt} ($\sqrt{s}=7$~{\rm TeV}) are
  compared to the NNLO calculation~\cite{Catani:2011qz}.}
\label{fig:deltaPHI} 
\end{figure} 
\nin
%%%%%%%%%%%%%%%%%%%%%%%
%In particular,
%the impact of the full NNLO corrections turns out to be reasonably moderate, 
%with a K-factor, defined as the ratio between the NNLO and NLO+box distributions,
%of about ${\rm K}\simeq 1.35$. 
We find that about 30\% of the NNLO corrections is due to the $gg$ channel (the 
{\it box contribution} is responsible for more than half of it), while almost
60\% still arises from the next-order corrections to the $qg$ channel.
The NNLO calculation includes the perturbative corrections from the entire phase
space region 
(in particular, the next-order correction to the dominant $qg$ channel)
and the contributions from all possible partonic channels (in particular,
a fully-consistent treatment of the {\it box contribution} to 
the $gg$ channel~\footnote{The calculation \cite{Bern:2002jx} of the next-order
gluonic corrections to the  {\it box contribution} indicates an increase of 
the NNLO result by less than 10\% if $M_{\gamma\gamma} \geq 100$~GeV.}).
Owing to these reasons, the NNLO result can be considered a reliable estimate of
direct diphoton production, although further studies (including independent
variations of $\mu_R$ and $\mu_F$, and detailed analyses of kinematical distributions)
are necessary to quantify the NNLO theoretical uncertainty.

\vspace*{0.2cm}

\nin
Recent results from the LHC~\cite{Chatrchyan:2011qt,Aad:2011mh} and the
Tevatron \cite{Aaltonen:2011vk} show some discrepancies between the data
and NLO theoretical calculations of diphoton production. Basically, discrepancies were found in
kinematical regions where the NLO calculation is {\em effectively} a LO theoretical 
description of the process. Such 
phase space regions (away from the back-to-back configuration) are accessible at 
NLO for the first time, due to the final-state radiation of the 
additional parton~\footnote{The low-mass
region ($M_{\gamma\gamma}\leq 80GeV$) in Figure \ref{fig:mass} also belongs to this case.}. Figure~\ref{fig:deltaPHI} 
shows a measurement by CMS~\cite{Chatrchyan:2011qt}, of the diphoton
cross section as a function of the azimuthal angle $\Delta \phi_{\gamma\gamma}$ between the
photons. The data are compared with our NLO and NNLO calculations~\cite{Catani:2011qz}.
The acceptance criteria used in this analysis ($\sqrt{\rm s}=7$~TeV) require: $p_T^{\rm harder}\geq
23$~GeV and $p_T^{\rm softer}\geq 20$~GeV.
The rapidity of both photons is restricted to $|y_\gamma| \leq 2.5$, and
the invariant mass of the diphoton system is constrained to be $M_{\gamma\gamma} > 80\,{\rm GeV}$. The isolation parameters
have the values $\epsilon_\gamma=0.05$, $n=1$ and $R=0.4$. 

\vspace*{0.2cm}

\nin
The histograms in Fig.~\ref{fig:deltaPHI} show
that the NNLO QCD results remarkably improve the theoretical description of the CMS data throughout the 
entire range of $\Delta \phi_{\gamma\gamma}$.

\vspace*{0.2cm}

\begin{figure}[hbt] 
\centerline{\includegraphics[width=7.9cm,height=6.1cm]{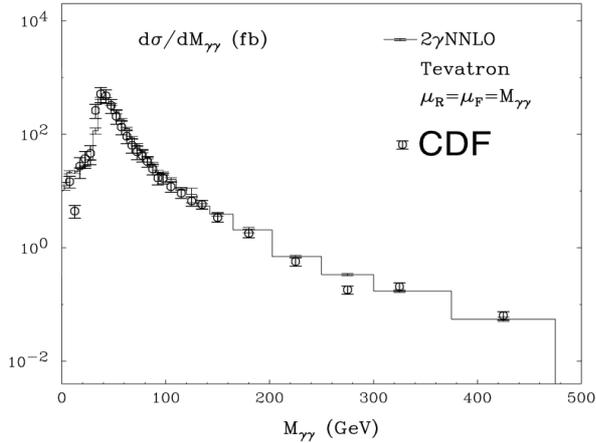}}
%{\epsfig{figure=mpsi2mc.eps,height=70mm}}
\caption{\scriptsize Diphoton 
cross section as a function of the invariant mass of
  the two photons. Data from CDF~\cite{Aaltonen:2011vk} ($\sqrt{s}=1.96$~{\rm TeV}) are
  compared to the NNLO calculation.}
\label{fig:tev} 
\end{figure}

\nin
In Figure~\ref{fig:tev}, we present the invariant mass distribution for 
diphoton production at the Tevatron ($\sqrt{s}=1.96$~TeV) calculated at NNLO, compared with a measurement performed by CDF~\cite{Aaltonen:2011vk}. The acceptance criteria in this case require: $p_T^{\rm harder}\geq
17$~GeV and $p_T^{\rm softer}\geq 15$~GeV. The rapidity of both photons is restricted to $|y_\gamma| \leq 1$. The isolation parameters are set to the values $E_{T~max}=2$~GeV, $n=1$ and $R=0.4$, and the minimum angular separation between the two photons is $R_{\gamma\gamma}=0.4$.
%%%%%%%%%%%%%%%%%%%%%%%
\nin
%%%%%%%%%%%%%%%%%%%%%%%
Though in this case the increase from the LO to the NLO result is considerably smaller than at the LHC~\cite{Catani:2011qz}, the NNLO QCD corrections still improve remarkably the theoretical description of the CDF data, in particular in the low mass region ($M_{\gamma\gamma} \leq 2p_T^{harder}=34$~GeV).
%\vspace*{0.2cm}

\medskip

\nin
We note that the CMS and CDF data are obtained by using the standard cone isolation criterion and the constraint in Eq.~(\ref{eq:etmax}) is applied only to the cone of radius $r=R$. Since the smooth isolation criterion used in our calculation (we apply Eq.~(\ref{eq:etmax}) for all cones with $r\leq R$) is stronger than the photon
isolation used by CMS and CDF, we remark that our NLO and NNLO results cannot overestimate the corresponding
theoretical results for the experimental isolation criterion. 
%The NNLO result sizeably increases the NLO result.
%Owing to transverse momentum conservation, the photons produced by the LO subprocess 
%$q\bar{q}\rightarrow \gamma\gamma$ have $\Delta \phi_{\gamma\gamma}=\pi$ and, therefore, the LO calculation
%gives a vanishing contribution to the histogram bins in the Figure.
%This Figure illustrates the fact that even NLO calculations may fail dramatically in some regions
%of phase space.
%
%In this case, the problem arises because the LO cross section of diphoton production 
%is zero in the region of $\Delta \phi_{\gamma\gamma} < \pi$, a simple consequence
%of momentum conservation, and thus NLO is actually the lowest non-zero order.
%In this particular instance, the problem arises because in the region
%of $\Delta \phi_{\gamma\gamma} < \pi$ the cross section is zero in
%leading-order diphoton production, a simple consequence of momentum
%conservation.
%
%Thus NLO is actually the lowest non-zero order. 
%Since NNLO is effectively NLO and, additionally, 
%introduces new topologies (e.g. $qq \to qq\gamma \gamma$), 
%the cross section receives large corrections and we obtain, therefore, much better agreement with the data.

\vspace*{0.8cm}

\nin
The results illustrated in this contribution show that the NNLO description of diphoton 
production is essential to understand the 
phenomenology associated to this process, and therefore, the NNLO calculation is a relevant tool 
to describe the main background for Higgs boson searches and studies.
%%%%%%%%%%%%%%%%%%%%%%%%%%%
\section*{Acknowledgements}
\nin
%%%%%%%%%%%%%%%%
I would like to thank Stefano Catani and Daniel de Florian for helpful comments. This work was 
supported by the INFN and the Research Executive Agency (REA) of the European Union under the Grant Agreement number PITN-GA-2010-264564 (LHCPhenoNet). 


%% References


%\bibliographystyle{elsarticle-num}





% \end{thebibliography}

%%%%%%%%%%%%%%%%%%%%
%\vfill\eject

%%%%%%%%%%%%%%%

\end{document}